\documentclass[         %
aps,                    
prd,                    
showpacs,               
superscriptaddress,     
nofootinbib,            
twocolumn,              %
showkeys,               %
preprintnumbers,        %
amsmath,                %
amssymb,                %
floatfix]               
{revtex4-1}             
\usepackage{graphicx, amsmath}
\graphicspath{{figures/},{figures_yamac/}}
\usepackage{mathrsfs, bm} 
\usepackage{mathtools} 
\usepackage{amsmath}
\usepackage{amssymb}
\usepackage[utf8]{inputenc}
\usepackage{multirow}
\usepackage{color}
\usepackage{hyperref}
\usepackage{nameref}
\usepackage{nicefrac}
\usepackage{soul}
\usepackage{microtype} 
\usepackage{mhchem}
\usepackage[dvipsnames]{xcolor}

\begin{document}
\title{Time dependent signatures of core-collapse supernova neutrinos at HALO}
\author{B. Ekinci}
\email{Corresponding author: baekinci@gmail.com}
\affiliation{Mimar Sinan Fine Arts University, Sisli, Istanbul, 34380, Turkey}
\author{Y.~Pehlivan}
\email{yamac.pehlivan@msgsu.edu.tr}
\affiliation{Mimar Sinan Fine Arts University, Sisli, Istanbul, 34380, Turkey}
\author{Amol V. Patwardhan}
\email{apatward@slac.stanford.edu}
\affiliation{SLAC National Accelerator Laboratory, 2575 Sand Hill Road, Menlo Park, CA, 94025}
\date{\today}
\begin{abstract}
We calculate the response of a lead-based detector, such as the Helium and Lead
Observatory (HALO) or its planned upgrade HALO-1kt to a galactic core-collapse
supernova. We pay particular attention to the time dependence of the reaction
rates. All reaction rates decrease as the neutrino luminosity exponentially
drops during the cooling period but the ratio of one-neutron (1n) to two-neutron
(2n) event rates in HALO is independent of this overall decrease. Nevertheless,
we find that this ratio still changes with time due to the changing character of
neutrino flavor transformations with the evolving conditions in the supernova.
In the case of inverted hierarchy, this is caused by the fact that the spectral
splits become less and less sharp with the decreasing luminosity. In the case of
normal hierarchy, it is caused by the passage of the shock wave through the
Mikheyev-Smirnov-Wolfenstein resonance region. However, in both cases, we find
that the change in the ratio of 1n to 2n event rates is limited to a few
percent. 
\end{abstract}
\medskip

\preprint{SLAC-PUB-17582, N3AS-21-002}

\preprint{}
\maketitle

\section{Introduction}
\label{section:introduction}

Based on statistical and observational studies, somewhere between $0.5-3$
supernova explosions should happen in our galaxy per century
\citep{Cappellaro:1999qy, Diehl:2006cf}. The latest supernova exploded about
$120$ years ago \cite{Reynolds:2008if} but it was not optically detected because
it was buried in dust near the galactic center. However, with today's neutrino
detection capabilities, the next galactic core-collapse supernova will leave
thousands of events in the neutrino detectors
\cite{Scholberg:2012id,Nagakura:2020bbw} even if it explodes near the center of
the galaxy. The question of what we can learn about the core-collapse supernovae
and about neutrinos themselves from such a signal is a prevailing one. A recent
review can be found in Ref.  \cite{Horiuchi:2017sku}. 

In this paper we are interested in calculating the response of a lead-based
detector such as HALO or HALO-1kt to a galactic core-collapse
supernova.\footnote{From now on to be referred simply as a supernova.} Neutrinos
and antineutrinos of all flavors are emitted from a supernova \cite{barkat,
Woosley:2002zz}. These neutrinos undergo flavor evolution after they thermally
decouple from the proto-neutron star at the center. Besides the ordinary vacuum
oscillations, this involves collective neutrino oscillations which happen due to
coherent neutrino-neutrino scattering in the inner regions
\citep{Pantaleone:1992xh, Pantaleone:1992eq, Duan:2010bg, Chakraborty:2016yeg},
and Mikheyev-Smirnov-Wolfenstein (MSW) resonances due to coherent scattering on
background electrons \cite{Wolfenstein:1977ue, Mikheev:1986wj, Kuo:1989qe} in
the outer regions.  HALO can detect these neutrinos through charged-current (CC)
reactions
\begin{equation}
\label{CC}
\begin{split}
\nu_e + \ce{^{208}Pb} &\longrightarrow \ce{^{207}Bi} + n + e^-,\\
\nu_e + \ce{^{208}Pb} &\longrightarrow \ce{^{206}Bi} + 2n + e^-,\\
\end{split}
\end{equation}
and neutral-current (NC) reactions
\begin{equation}
\label{NC}
\begin{split}
\overset{(-)}{\nu} + \ce{^{208}Pb} &\longrightarrow \ce{^{207}Pb} + n + \overset{(-)}{\nu}, \\
\overset{(-)}{\nu} + \ce{^{208}Pb} &\longrightarrow \ce{^{206}Pb} + 2n +\overset{(-)}{\nu},
\end{split}
\end{equation}
on $79$ tons of \ce{^{208}_{82}Pb} target \cite{ZUBER2015233}. HALO-1kt will use
$1$ kiloton of lead target and will have significantly increased efficiency.
All neutrino and antineutrino flavors can participate in NC reactions as
indicated by the ${(-)}$ overset in Eq. (\ref{NC}). As a result, these reactions
are insensitive to flavor transformations. But the CC reactions can only be
participated by electron neutrinos. Since $\ce{^{208}Pb}$ is a neutron-rich
nucleus, reactions with electron antineutrinos are suppressed by Pauli blocking.
As a result, HALO is primarily sensitive to the flavor transformations of
electron neutrinos inside the supernova. Currently, neutrino-lead cross sections
are not experimentally known. But theoretical calculations are carried out by
several groups \cite{Kolbe:2000np, Engel:2002hg, Lazauskas:2007bs,
Almosly:2016mse, Almosly:2019han, Ejiri:2019ezh}. Estimates of expected event
rates at HALO are based on such calculations
\cite{Rosso:2020owy,Vaananen:2011bf, Vale:2015pca, Bandyopadhyay:2016gkv}. 

The first detailed study of the HALO event rates which took neutrino flavor
evolution into account was Ref. \cite{Vaananen:2011bf}. The primarily interest
of this study was to extract information about the neutrino energy spectra
emitted during the cooling period of the supernova from the HALO signal. To be
able to scan the large parameter space, the authors adopted an approach which is
based on modifying the neutrino energy spectra manually to mimic the dynamical
flavor evolution. In particular, based on the results from earlier flavor
evolution studies they assumed that sharp spectral swaps \cite{Duan:2006an,
Duan:2006jv, Raffelt:2007cb, Raffelt:2007xt, Duan:2007bt, Dasgupta:2010cd,
Choubey:2010up} occur in fixed parts of the neutrino energy spectra due to
collective neutrino oscillations. They also ignored the effects of the shock
wave. This practical approach allowed them to write down analytical expressions
for the neutrino spectra reaching Earth for a large variety of initial
conditions, and show that HALO signal can be used to extract information about
the spectral pinching parameters. A similar approach was adopted in Ref.
\cite{Vale:2015pca} which was concerned with determining the neutrino mass
hierarchy by combining the electron neutrino signal from a lead or iron detector
and the electron antineutrino signal from a water-Cherenkov detector. A
complementary study was carried out more recently in Ref.
\cite{Bandyopadhyay:2016gkv}, in which the authors used time dependent neutrino
luminosities and initial energy distributions taken from a proto-neutron star
simulation \cite{Fischer:2009af}. Since this simulation produced very similar
fluxes for different flavors, and their primary purpose was to compare lead vs
iron-based detectors in terms of total number of events, the authors ignored the
collective neutrino oscillations and the effects shock wave.

This paper aims to estimate the event rates in HALO during a galactic supernova
based on a fully \emph{dynamical} and \emph{time dependent} calculation of
neutrino flavor evolution. Here, being \emph{dynamical} refers to the explicit
solutions of flavor evolution equations starting from the surface of the
proto-neutron star and ending on the surface of the supernova. Collective
oscillations and MSW resonances naturally appear in these dynamical solutions.
The \emph{time dependence} mentioned above refers to the fact that we include
(i) the decreasing neutrino luminosity of the proto-neutron star during the
cooling period and (ii) the propagating shock wave in the mantle. 

During the cooling period, neutrino luminosities are expected to drop roughly as
$L_{\nu_\alpha} e^{-t/\tau}$ \cite{Mirizzi:2015eza, Roberts:2016rsf}. Here, $t$
denotes the post-bounce time, $\tau$ is the relevant time scale, $\nu_\alpha$
with $\alpha=e,\mu,\tau$ denotes the neutrino flavor, and $L_{\nu_\alpha}$ is
the initial $\nu_\alpha$ luminosity. This will lead to exponentially decreasing
reaction rates in all detectors. However, there is a more subtle effect: with
decreasing neutrino luminosity, collective neutrino oscillations change their
character, especially in the case of inverted hierarchy (IH). Our calculations
for IH indicate that initially, when the neutrino luminosity is large, the
spectral swaps are sharp. In other words, completely adiabatic or diabatic
transitions occur across the whole spectrum, depending on the energy. As the
neutrino luminosity drops, we see partially adiabatic transitions in the part of
the spectrum above $35$ MeV. This is the part of the spectrum to which a
lead-based detector is most sensitive, i.e., most of the neutrons produced in
the detector will originate from neutrinos in this energy region
\cite{Bandyopadhyay:2016gkv}. Therefore, it is reasonable to ask whether this
effect can lead to a time dependence beyond the exponential drop in reaction
rates in the case of IH. In the case of normal hierarchy (NH), collective
oscillations affect only the very low energy part of the spectrum. For this
reason, one does not expect to see a similar effect for NH. 

The second source of time dependence in the HALO signal is the propagating shock
wave. This effect can be expected to appear when the shock wave reaches the MSW
resonance region. There are two MSW resonances: The low resonance occurs in the
outermost layers of the mantle and is experienced by neutrinos in both NH and
IH. However, one does not expect to see its effect in reaction rates because the
shock wave reaches there at very late times when the neutrino luminosity drops
to a few percent of its initial value. The high resonance occurs at relatively
higher densities and is experienced by neutrinos only in the case of NH. The
shock wave reaches this region relatively earlier, while there is still a
sizable neutrino luminosity. Therefore, in principle, it is possible to see the
effects of the shock wave in reaction rates in the case of NH. 

The impact of the supernova shock waves on neutrino flavor evolution has been
examined in several works over the years \cite{Schirato:2002tg, Fogli:2003dw,
Takahashi:2002yj, Tomas:2004gr, Dasgupta:2005wn, Friedland:2006ta, Gava:2009pj,
Friedland:2020ecy}.  The combined effect of the collective neutrino oscillations
and the shock wave modification of MSW resonances was first considered in Ref.
\cite{Gava:2009pj}. In this work, an interesting interplay between the two
features  was pointed out for electron antineutrinos which feel both effects in
the case of IH. This has observational consequences for  water-Cherenkov and
liquid scintillator experiments.  However, no such interplay is to be expected
in the HALO signal: HALO is sensitive to the flavor evolution of electron
neutrinos only, and they are affected either by collective oscillations in the
case of IH, or by the shock wave modification of MSW resonances in the case of
NH. This separation of effects is one of the motivations of the present study. 

Another motivation is the fact that HALO has an adequate timing mechanism to
look for dynamical effects in a supernova. It has a time resolution of about
$200$ $\mu$s, which is the time required for the emitted neutrons to thermalize
before being captured at the \ce{^3 He} counters
\cite{2018npa..confE.474P,Virtue}. In addition, all events are time stamped by a
GPS-based system \cite{Virtue}, allowing a comparison with other neutrino
observatories.

Our treatment covers neither the dynamical aspects of neutrino flavor evolution
nor the time dependent features of the supernova neutrino signal fully. As for
the dynamical flavor evolution, we leave out the multiangle nature of the
collective neutrino oscillations: we treat the dependence of the
neutrino-neutrino interactions on the angle between the neutrinos
\cite{Sigl:1992fn} in an effective way using the so-called \emph{neutrino bulb
model} \cite{Duan:2006an}.  The multiangle effects are expected to delay the
onset of collective oscillations but leave its general features intact for the
type of initial spectra that we consider \cite{Mirizzi:2010uz}. We also do not
take the time evolution of the proto-neutron star itself into account as is
done, for example, in Ref. \cite{Bandyopadhyay:2016gkv}. In reality, the radius
of the proto-neutron star, and the energy distributions of the emitted neutrinos
slightly change during the cooling period. See, for example, Refs.
\cite{Mirizzi:2015eza, Roberts:2016rsf}. Our purpose in leaving out this
important feature is to isolate and examine the other time dependent features
mentioned above. 

We start in Section II with a general description of the neutrino flavor
evolution equations in the supernova. In Section III, we provide a
semi-analytical view of flavor evolution with an emphasis on the sharpness of
spectral splits in terms of a simple jumping probability $p$ between the matter
eigenstates in the collective oscillation region. We present flavor evolution
examples and show the departure from the sharp spectral splits in the high
energy region. The discussion in this section is useful in interpreting our
results in the case of IH. In Section IV, we calculate the expected reaction
rates per kiloton of lead target for four different neutrino energy distribution
models as a function of time. In particular we present the results for the time
dependence of the ratio of 1n to 2n events. We present our conclusions in
Section V. 

\section{Supernova Neutrinos}

We assume that the supernova is spherically symmetric, and that neutrinos of all
flavors and energies thermally decouple at a single sharp surface close to the
surface of the proto-neutron star. After that, neutrinos freely stream uniformly
in all directions. This is known as the \emph{neutrino bulb model}, which was
developed in Ref. \cite{Duan:2006an}. We also adopt the single-angle
approximation in describing the neutrino flavor evolution. Both of these
approaches are summarized below.
\begin{figure}
\begin{center}
\includegraphics[clip=true,trim={12cm 1cm 2cm 1cm},width=0.6\columnwidth]{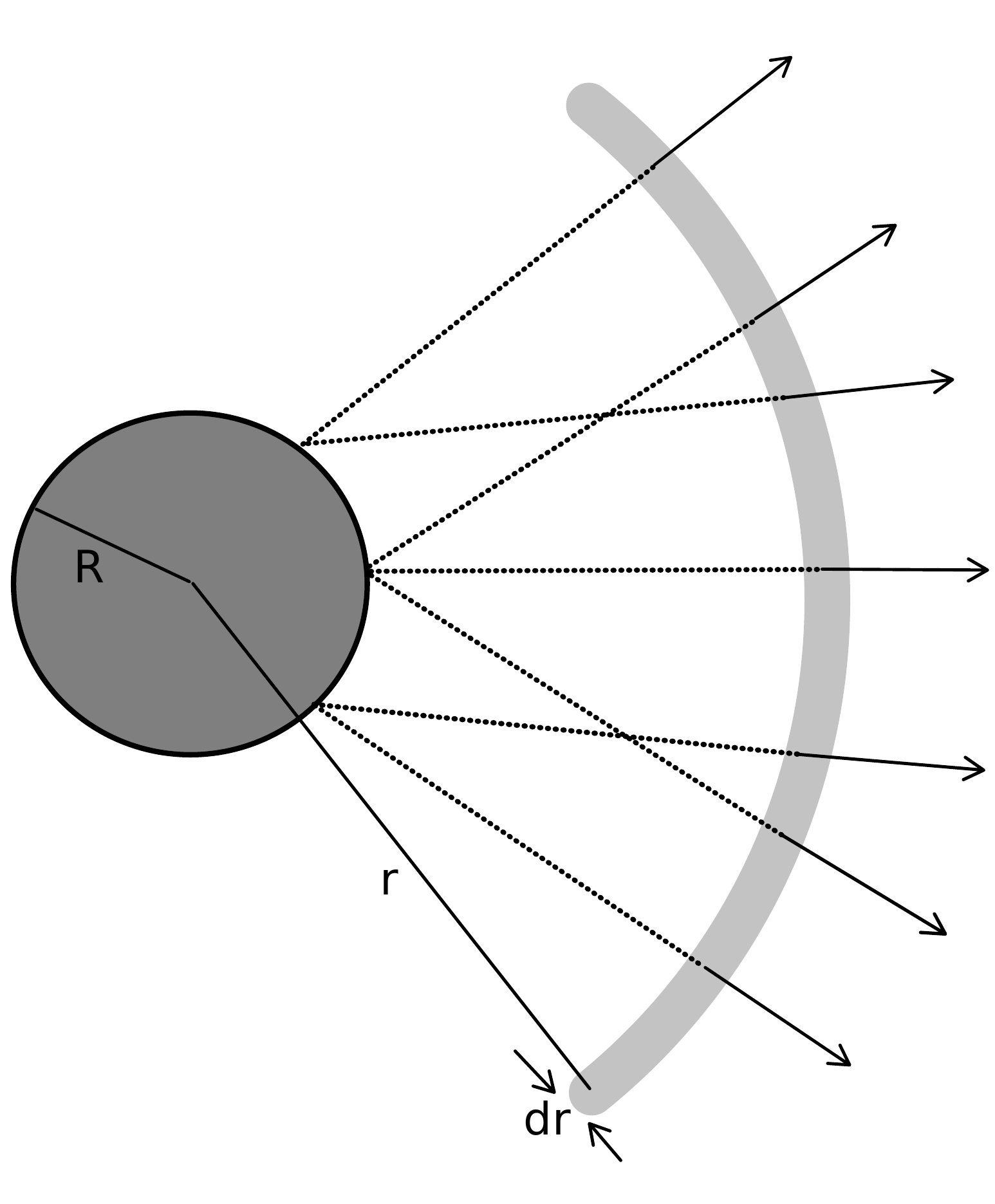}
\caption{Schematic picture of neutrino emission from the proto-neutron star and the
neutrino ensemble represented by the density operator $\hat{\rho}_t(E,r)dtdE$ in the
neutrino bulb model \cite{Duan:2006an}. See the text for details.}
\label{bulb_model_yamac}
\end{center}
\end{figure}

Let us consider all neutrinos with energy between $E$ and $E+dE$ which are
emitted from the surface of the proto-neutron star during a post-bounce time
interval between $t$ and $t+dt$. Not all of these neutrinos travel radially
because emission can happen in any angle from the surface of the proto-neutron
star. (See Fig. \ref{bulb_model_yamac}.) But, in the single angle approximation,
it is assumed that these neutrinos reach a given distance (say $r$) from the
center of the supernova at the same time. In other words, small differences in
their paths are omitted. Furthermore, the single angle approximation and the
spherical symmetry assumption together imply that any two neutrinos in this
group evolve in the same way regardless of their direction, provided that they
are initially emitted in the same flavor. Therefore, these neutrinos form an
ensemble which evolves with distance $r$. Here we choose to work with
\emph{unnormalized} density operators, i.e., we want to define an infinitesimal
density operator $\hat{\rho}_t(E,r)dtdE$ whose trace is equal to the
\emph{total} number of neutrinos with energy between $E$ and $E+dE$, emitted
from the proto-neutron star between $t$ and $t+dt$. These neutrinos occupy the
shaded region between $r$ and $r+dr$ shown in Fig. \ref{bulb_model_yamac} at a
later time. Here, we use the natural units so that $dr=dt$. 

On the surface of the proto-neutron star where $r=R$, this density operator is given by  
\begin{equation}
\label{initial rho}
\hat \rho_t(E,R)=\sum_{\alpha=e,\mu,\tau} \frac{L_{\nu_\alpha} e^{-t/\tau}}{\langle E_{\nu_{\alpha}}\rangle} 
f_{\nu_\alpha}(E) |\nu_\alpha\rangle\langle \nu_\alpha|.
\end{equation}
The above equation is easy to understand: $L_{\nu_\alpha}e^{-t/\tau}$ is the
total energy emitted per unit time in terms of $\nu_\alpha$'s. Dividing this by
their average energy $\langle E_{\nu_{\alpha}}\rangle$, we find the number of
$\nu_\alpha$'s emitted per unit time. Finally, multiplication with their
normalized energy distribution $f_{\nu_\alpha}(E)$ yields the number of those
$\nu_\alpha$'s per unit energy interval. For antineutrinos, we define an
analogous expression $\hat{\bar\rho}_t(E,R)$ which is the same as in Eq.
(\ref{initial rho}) except that neutrino quantities are replaced with
antineutrino quantities labeled by $\bar\nu_\alpha$, e.g., $L_{\bar\nu_\alpha}$,
$\langle E_{\bar\nu_{\alpha}}\rangle$, etc $\dots$ 

The neutrino luminosities used in Eq. (\ref{initial rho}) can be found from the
total binding energy $E_B$ emitted by the supernova. Since almost all of this
energy is emitted in terms of neutrinos during the cooling period, divided
equally between all neutrino and antineutrino flavors, neutrino luminosities can
be calculated from $E_B = 6 \times \int_0^\infty L_{\nu_\alpha} e^{-t/\tau} dt$.
In our simulations, we use $\tau=3$ s \cite{Mirizzi:2015eza, Roberts:2016rsf}.
We take the total released gravitational binding energy as $E_B=5.72\times
10^{53}$ ergs, which is close to the upper limit\footnote{Note that the observed
mass of the residual neutron star mass suggests a more conservative range of
$1.0 \times 10^{53} \mbox{ergs} <E_B< 4.0 \times 10^{53} \mbox{ergs}$
\cite{doi:10.1146/annurev.aa.27.090189.003213}. Also see the discussion in Ref.
\cite{Yoshida:2003ay}.} of the binding energy released by the SN1987A as
calculated from the observed $\bar\nu_e$ signal \cite{Suzuki:1994qp}. This leads
to the initial neutrino luminosities given by $L_{\nu_\alpha}=L_{\bar\nu_\alpha}
= 3.18\times 10^{52}$ erg/s, which drops exponentially after that.
 
\begin{table}
\begin{center}
\begin{tabular}{ | l l |}
\hline 
\multicolumn{2}{|c|}{Supernova}  \\ 
\hline 
\multicolumn{2}{|l|}{Initial neutrino luminosities: $L_{\nu_\alpha}=L_{\bar\nu_\alpha}=3.18\times 10^{52}$ erg/s }\\ 
\multicolumn{2}{|l|}{Decay time $\tau=3.0$ s}\\ 
\multicolumn{2}{|l|}{Total energy released: $E_B=5.72\times 10^{53}$ erg }\\ 
\multicolumn{2}{|l|}{Neutron star radius: $R=10$ km}\\ 
\multicolumn{2}{|l|}{Distance to Earth: $d=10$ kpc}\\ 
\multicolumn{2}{|l|}{Average energies}  \\ 
\multicolumn{2}{|l|}{Model I:
$\langle E_{\nu_{e}}\rangle=8 \mbox{ MeV} \;\; 
\langle E_{\bar\nu_{e}}\rangle=11 \mbox{ MeV}  \;\;
\langle E_{\nu_x}\rangle=13 \mbox{ MeV}$} 
\\
\multicolumn{2}{|l|}{Model II:
$\langle E_{\nu_{e}}\rangle=8 \mbox{ MeV} \;\; 
\langle E_{\bar\nu_{e}}\rangle=11 \mbox{ MeV}  \;\;
\langle E_{\nu_x}\rangle=16 \mbox{ MeV}$} 
\\
\multicolumn{2}{|l|}{Model III:
$\langle E_{\nu_{e}}\rangle=8 \mbox{ MeV} \;\; 
\langle E_{\bar\nu_{e}}\rangle=11 \mbox{ MeV}  \;\;
\langle E_{\nu_x}\rangle=20 \mbox{ MeV}$} 
\\
\multicolumn{2}{|l|}{Model IV:
$\langle E_{\nu_{e}}\rangle=15 \mbox{ MeV} \;\; 
\langle E_{\bar\nu_{e}}\rangle=20 \mbox{ MeV}  \;\;
\langle E_{\nu_x}\rangle=25 \mbox{ MeV}$} 
\\
\hline 
\multicolumn{2}{|c|}{Neutrino Mixing}  \\ 
\hline 
\multicolumn{2}{|l|}{$\sin\theta_{12}=0.554 \;\; \sin\theta_{13}=0.148 \;\;
m_2^2-m_1^2=7.53\times10^{-5} \mbox{eV}^2$}\\
\multicolumn{2}{|l|}{$\sin\theta_{23}=0.715 \;\; m_3^2-m_2^2=2.44 \times 10^{-3} \mbox{eV}^2$ (for NH)} \\ 
\multicolumn{2}{|l|}{$\sin\theta_{23}=0.732 \;\; m_3^2-m_2^2=-2.55 \times 10^{-3} \mbox{eV}^2$ (for IH)} \\ 
\hline 
\multicolumn{2}{|c|}{Detector}  \\ 
\hline 
\multicolumn{2}{|l|}{Mass = $1$ kt}\\
\multicolumn{2}{|l|}{Reactions and threshold energies}\\
$\nu_e + \ce{^{208}Pb} \longrightarrow \ce{^{207}Bi} + n + e^-$ & $E_{\mbox{\tiny th}}=9.76$ MeV\\
$\nu_e + \ce{^{208}Pb} \longrightarrow \ce{^{206}Bi} + 2n + e^- $ & $E_{\mbox{\tiny th}}=17.86$ MeV\\
$\overset{(-)}{\nu} + \ce{^{208}Pb} \longrightarrow \ce{^{207}Pb} + n + \overset{(-)}{\nu}$ & $E_{\mbox{\tiny th}}=7.37$ MeV\\
$\overset{(-)}{\nu} + \ce{^{208}Pb} \longrightarrow \ce{^{206}Pb} + 2n + \overset{(-)}{\nu}$ & $E_{\mbox{\tiny th}}=14.11$ MeV\\
\hline 
\end{tabular}
\caption{The summary of the parameters that we use in our numerical calculations. } 
\label{parameters} 
\end{center}
\end{table}

Unlike the luminosities, the energy distributions of different neutrino and
antineutrino flavors are not the same because they are subject to different
interactions inside the proto-neutron star. Here we adopt the fit function
provided by Ref. \cite{2003ApJ...590..971K} by setting their dimensionless
fitting parameter to $3$ and normalizing accordingly, which gives 
\begin{equation}
\label{equ4}
f_{\nu_{\alpha}}(E)=\frac{128E^3}{3\langle E_{\nu_{\alpha}} \rangle^4} 
\exp\left(-\frac{4E}{\langle E_{\nu_{\alpha}} \rangle}\right).
\end{equation}
Simulations of neutrino transport in proto-neutron star generally lead to
average energies around $10$-$20$ MeV. (See Ref. \cite{Mathews:2014qba} for a
partial compilation of these simulations and the references therein.) In this
paper, we use four different sets of average energies which are referred to as
models I-IV in Table \ref{parameters}. We choose these particular sets of
average energies in order to cover qualitatively different scenarios as
discussed in the next section. 

As the neutrinos move outward, the density operator representing our ensemble
changes according to   
\begin{equation}
\label{equ1}
\frac{d}{d r} \hat\rho_t(E, r)=-i[\hat H_t(E,r), \hat\rho_t(E, r)]. 
\end{equation}
Here we use $r$ as the proper time of the neutrinos because they propagate
almost with the speed of light. As they quickly fly through the supernova, they
essentially see the background as it is when they are emitted at time $t$. This
is because most of the non-trivial flavor evolution (collective effects and the
MSW resonances) occurs before $10^5$ km, and it takes only a fraction of a
second for the neutrinos to reach there. For this reason, we use the Hamiltonian
$H_t(E,r)$ which represents the conditions in the supernova at time $t$ to
describe the evolution of $\hat\rho_t(E,r)$. This Hamiltonian is given by 
\begin{equation}
\label{hamiltonian}
\begin{split}
\hat H_t(E,r)=& \sum_i \frac{m_i^2}{2E}|\nu_i\rangle\langle\nu_i| +
\sqrt{2}G_F N_e(t,r)|\nu_e\rangle\langle\nu_e| 
\\ +&\frac{\sqrt{2} G_{\mathrm{F}}}{2 \pi R^{2}} D(\tfrac{R}{r})
\int (\hat\rho_t(E', r) - \hat{\bar\rho}_t(E', r))dE'. 
\end{split}
\end{equation}

Here, the first term represents the vacuum oscillations. $\nu_i$ is the
eigenstate of mass $m_i$ for $i=1,2,3$, which is related to the flavor
eigenstates via $\langle \nu_\alpha|\nu_i\rangle=U_{\alpha i}$ where $U$ is the
neutrino mixing matrix \cite{Zyla:2020zbs}. Subtracting the trace of this term
from the Hamiltonian allows one to use only mass squared differences given in
Table \ref{parameters}. 

\begin{figure}
\begin{center}
\includegraphics[clip=true,trim={0cm 0cm 0cm 1.35cm},width=\columnwidth]{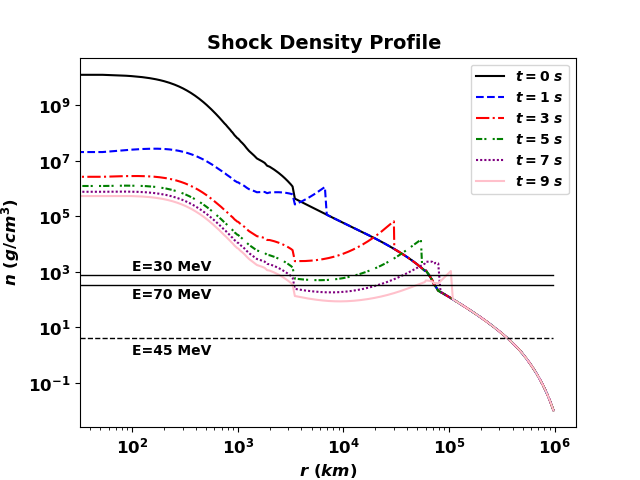}
\caption{The density for the progenitor model and various post-bounce times.
The solid horizontal lines represent the high resonance densities for $30$ MeV
and $70$ MeV neutrinos. The dashed horizontal line represents the low resonance
density for a $45$ MeV neutrino.}
\label{fig:density}
\end{center}
\end{figure}

The second term in Eq. (\ref{hamiltonian}) represents the CC forward scattering
of neutrinos from the background electrons and protons \cite{Wolfenstein:1977ue,
PhysRevD.22.2718} with $G_F$ denoting the Fermi interaction constant and
$N_e(t,r)$ denoting the net electron number density at radius $r$ at time $t$.
Assuming that the number of electrons per baryon is equal to $0.5$, electron
number density $N_e(t,r)$ is related to the mass density $n(t,r)$ by
\begin{equation}
N_e(t,r)=\left(\frac{n(t,r)}{10^8\mbox{g/cm}^3}\right)\; 3.01\times10^{31}
\mbox{cm}^{-3}.
\end{equation}
At $t=0$, we use the mass density profile provided by Ref.
\cite{1987ESOC...26..325N} as a $6 M_\odot$ helium-core progenitor model for
SN1987A. This profile is shown in Fig. \ref{fig:density} with the solid black
line. For post-bounce times, we obtain the density profile by superimposing a
parametric shock wave on the progenitor profile as described in Ref.
\cite{Fogli:2003dw}. The resulting post-bounce density profiles for $t=1,3,5,7$
s are also shown in Fig. \ref{fig:density}.

The last term in Eq. (\ref{hamiltonian}) represents the NC forward and exchange
scatterings of neutrinos from each other \cite{fuller&mayle, savage&malaney,
Pantaleone:1992eq, Pantaleone:1992xh, Sigl:1992fn}. The energy integral contains
all neutrinos at radius $r$. This may be confusing because it appears to imply
that every neutrino in the ensemble with energy $E$ meets and interacts with
every other neutrino in the whole system. However, this is not the case.
According to the neutrino bulb model, a \emph{test neutrino} interacts only with
those neutrinos which cross its path. But due to the assumption of spherical
symmetry and the single angle approximation, the evolution of the ones that it
meets is the same as those that it does not. As a result, one can use the
density operator of the whole system by simply multiplying it with a geometrical
factor which yields its relevant fraction. This geometrical factor is
$\tfrac{1}{2\pi R^{2}} D(\tfrac{R}{r})$ with \cite{Duan:2006an}
\begin{equation} \label{D}
D\left(\frac{r}{R}\right)=\frac{1}{2}\left[1-\sqrt{1-\left(\frac{R}{r}\right)^{2}}\right]^{2}.
\end{equation}
This factor also takes care of the angle dependence of the neutrino-neutrino
interactions in an effective way. The density operator $\hat{\bar\rho}_t(E, r)$
which appears the energy integral represents the antineutrinos in an analogous
way to neutrinos. Antineutrino density operators undergo the same evolution
except that their Hamiltonian should be obtained from Eq. (\ref{hamiltonian}) by
interchanging neutrino and antineutrino degrees of freedom and reversing the
sign of the CC interaction term. We do not consider any charge-parity violations
here because it would have no affect on the HALO signal.\footnote{This statement
is true as long as $\nu_\mu$, $\nu_\tau$, $\bar\nu_\mu$, $\bar\nu_\tau$ have the
same emission spectra \cite{Balantekin:2007es, Gava:2010kz, Kneller:2009vd,
Gava:2008rp, 2002PhLB..544..286Y}, and as long as the neutrino magnetic moment
is small \cite{Pehlivan:2014zua}.} 

\section{Flavor Evolution}

\subsection{Collective Oscillations}

It is helpful to describe the neutrino flavor evolution in the \emph{matter
basis}, which instantaneously diagonalizes the Hamiltonian. Since the
Hamiltonian changes with distance, so does the matter basis. We find it useful
to denote the matter basis at distance $r$ from the center of the supernova with
$|r_i\rangle$ where $i=1,2,3$ orders the matter eigenvalues from the lightest to
the heaviest. In other words, we write the Hamiltonian given in Eq.
(\ref{hamiltonian}) as  
\begin{equation}
\label{H in matter basis}
H(r)=\sum_{i=1}^3 \mathcal{E}_i(r)\; |r_i \rangle \langle r_i|,
\end{equation}
where $\mathcal{E}_1(r)<\mathcal{E}_2(r)<\mathcal{E}_3(r)$ is satisfied. For
simplicity, we drop the time and energy dependence of the Hamiltonian and the
density operator from our notation in the rest of the paper. Since the
Hamiltonian depends on the post-bounce time and the energy of the neutrino, so
do its eigenvalues and eigenstates on the right hand side of Eq. (\ref{H in
matter basis}). But this dependence is similarly suppressed in our notation.

For the models that we consider, we have 
\begin{equation}
\label{initial eigenbasis}
\begin{split}
|R_1\rangle &\approx \sin\theta_{23}|\nu_\mu\rangle + \cos\theta_{23}|\nu_\tau\rangle,  \\ 
|R_2\rangle &\approx \cos\theta_{23}|\nu_\mu\rangle - \sin\theta_{23}|\nu_\tau\rangle,  \\ 
|R_3\rangle &\approx |\nu_e\rangle, \quad  
\end{split}
\end{equation}
on the surface of the proto-neutron star where $r=R$. Since the luminosities are
equal and the $\nu_\mu$ and $\nu_\tau$ energy distributions are the same, the
initial density operator given in Eq. (\ref{initial rho}) can be written in the matter basis as 
\begin{figure}
\includegraphics[clip=true,trim={1cm 1cm 1cm 2cm},width=\columnwidth]{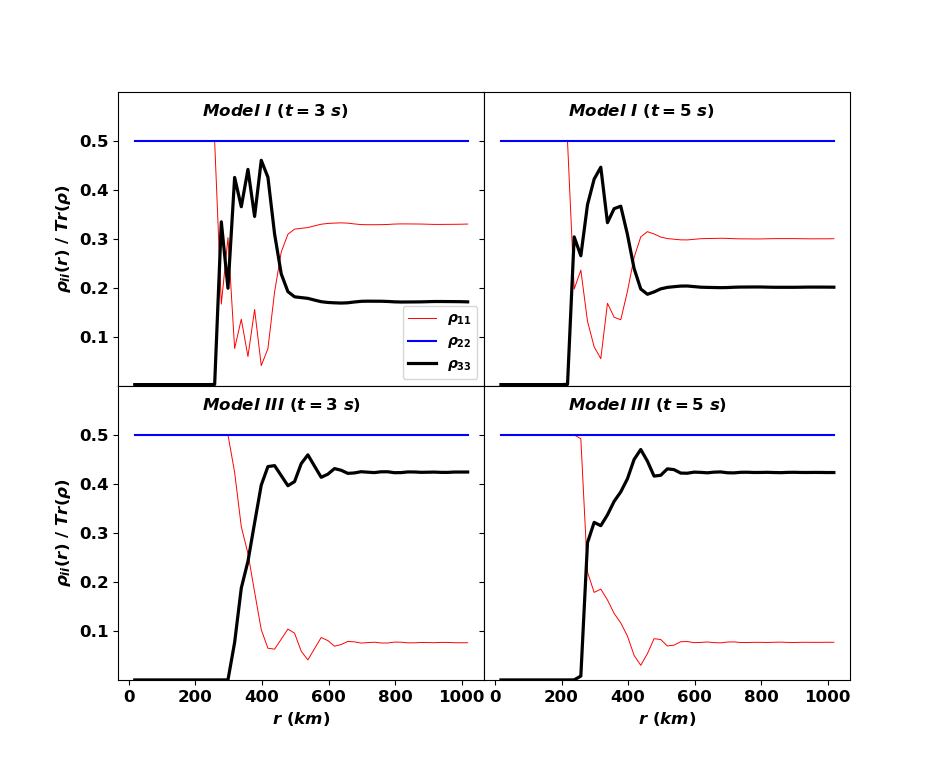}
\caption{The decoupling of the second matter eigenstate from the dynamics during
collective oscillations. The plot shows the diagonal elements of the density
operator in matter basis (i.e., $\rho_{ii}(r)=\langle
r_i|\hat\rho(r)|r_i\rangle$) divided by its trace. The thick black line
corresponds to the heaviest matter eigenstate which mixes with the lightest
shown with the thin red line. The blue line with medium thickness shows the
second matter eigenstate which decouples from the dynamics. This is for a $45$
MeV neutrino in IH for models I and III at $3$ s and $5$s. For other models,
energies, and times the results are similar.} 
\label{decoupling_of_2}
\end{figure}
\begin{equation}
\label{initial rho in matter}
\hat \rho(R)= \rho_{\mu\mu}(R)\left(| R_1 \rangle \langle R_1 | + | R_2 \rangle \langle R_2 |\right)
+ \rho_{ee}(R) | R_3 \rangle \langle R_3 |. 
\end{equation}
Here the flavor components $\rho_{\alpha\alpha}(R)$ are defined in Eq.
(\ref{initial rho}). For the models that we consider, collective oscillations
mix the first and third matter eigenstates while the second mass eigenstate
decouples from the dynamics. Examples of this behavior are shown in Fig.
\ref{decoupling_of_2}. This behavior is expected based on the results of earlier
studies, such as those in Refs. \cite{Choubey:2010up}. This tells us that $12$
and $23$ elements of the evolution operator in the matter basis are zero whereas
its $13$ component can be parametrized as
\begin{equation}
\label{transition probability}
\langle r_1| \mathcal{U}(r,R) | R_3 \rangle = \sqrt{p}e^{i\delta(r)}.
\end{equation}
Here, $\mathcal{U}(r,R)$ denotes the evolution operator from $R$ to $r$, and $p$
is the probability of $|R_3\rangle \leftrightarrow |r_1\rangle$ transition.
Both $p$ and the phase $\delta(r)$ depend on the neutrino energy and the
post-bounce time. But $p$ depends on the distance only for a short period during
the collective oscillations. Once the collective oscillations end at about $500$
km, $p$ is independent of distance (see Fig. \ref{decoupling_of_2}).  However
the phase $\delta(r)$ depends on the distance $r$ due to the
$\exp(-i\int\mathcal{E}_i(r)dr)$ terms picked up by the instantaneous
eigenstates during the evolution. As a result, after the collective oscillations
end the density operator is given by 
\begin{align}
\label{final rho in matter}
\hat \rho(r)&=
\left((1-p)\;\rho_{\mu\mu}(R)+p\;\rho_{ee}(R)\right) |r_1\rangle\langle r_1| \nonumber \\ 
& + \rho_{\mu\mu}(R)|r_2\rangle\langle r_2| \\
& + \left(p\;\rho_{\mu\mu}(R)+(1-p)\;\rho_{ee}(R)\right)|r_3\rangle\langle r_3| \nonumber \\ 
& +
\sqrt{p(1-p)}\left(\rho_{ee}(R)e^{i\delta(r)}-\rho_{\mu\mu}(R)e^{-i\delta(r)}\right)|r_1 \rangle\langle r_3| \nonumber\\
& + \mbox{h.c.} \nonumber
\end{align}
Here h.c. stands for hermitian conjugate of the $|r_1 \rangle\langle r_3|$ term.
The exponential phase terms cancel each other in the diagonal terms of the
density operator, but they do appear in the non-diagonal terms. 

Eq. (\ref{final rho in matter}) is true for both NH and IH, but with different
crossing probabilities. In the case of NH, the collective oscillations create a
spectral swap in the low energy part of the spectrum below a split energy
$E_{\mbox{\tiny NH}}$. In the case of IH, they create a spectral swap between a
low and a high split energy, denoted, respectively, by $E_{\mbox{\tiny IH}}$ and
$E_{\mbox{\tiny IH}}'$. In the ideal case of a \emph{sharp split}, $p$ is equal
to $1$ in the swapped region and $0$ in the unswapped region. In other words, in
the case of a sharp split one expects to find 
\begin{equation}
\label{p NH}
p\approx
\begin{cases}
1 & \mbox{for } E<E_{\mbox{\tiny NH}} \\
0 & \mbox{for } E>E_{\mbox{\tiny NH}}
\end{cases} \quad \mbox{ in NH,}
\end{equation}
and
\begin{equation}
\label{p IH}
p\approx
\begin{cases}
1 & \mbox{for } E_{\mbox{\tiny IH}}<E<E_{\mbox{\tiny IH}}'\\
0 & \mbox{ otherwise } 
\end{cases} \quad \mbox{ in IH. }
\end{equation}
NH spectral swap happens in the very low energy part of the spectrum. For all
the models that we considered, $E_{\mbox{\tiny NH}}$ was lower than all the
reaction threshold energies, which are listed in Table \ref{parameters}. But the
IH swap occurs in the region to which HALO is most sensitive. For the models
I-III, the low split energy $E_{\mbox{\tiny IH}}$ is around $7$-$9$ MeV. This is
close to, but lower than, the CC1n reaction threshold, which is $9.76$ MeV. In
order to see the effects of an IH low split energy which is higher than this
threshold, we set up the fourth model with higher average energies.  In model
IV, we find $E_{\mbox{\tiny IH}}=14$ MeV, which is higher than the CC1n reaction
threshold but lower than the CC2n reaction threshold. The high split energy
$E_{\mbox{\tiny IH}}'$ steadily increases as we go from model I to model IV. For
model I, it is $35$ MeV, and for model IV, it is $60$ MeV. It also increases
slightly with time within each model. 

\begin{figure}
\includegraphics[clip=true,trim={3cm 1cm 3cm 2cm},width=\columnwidth]{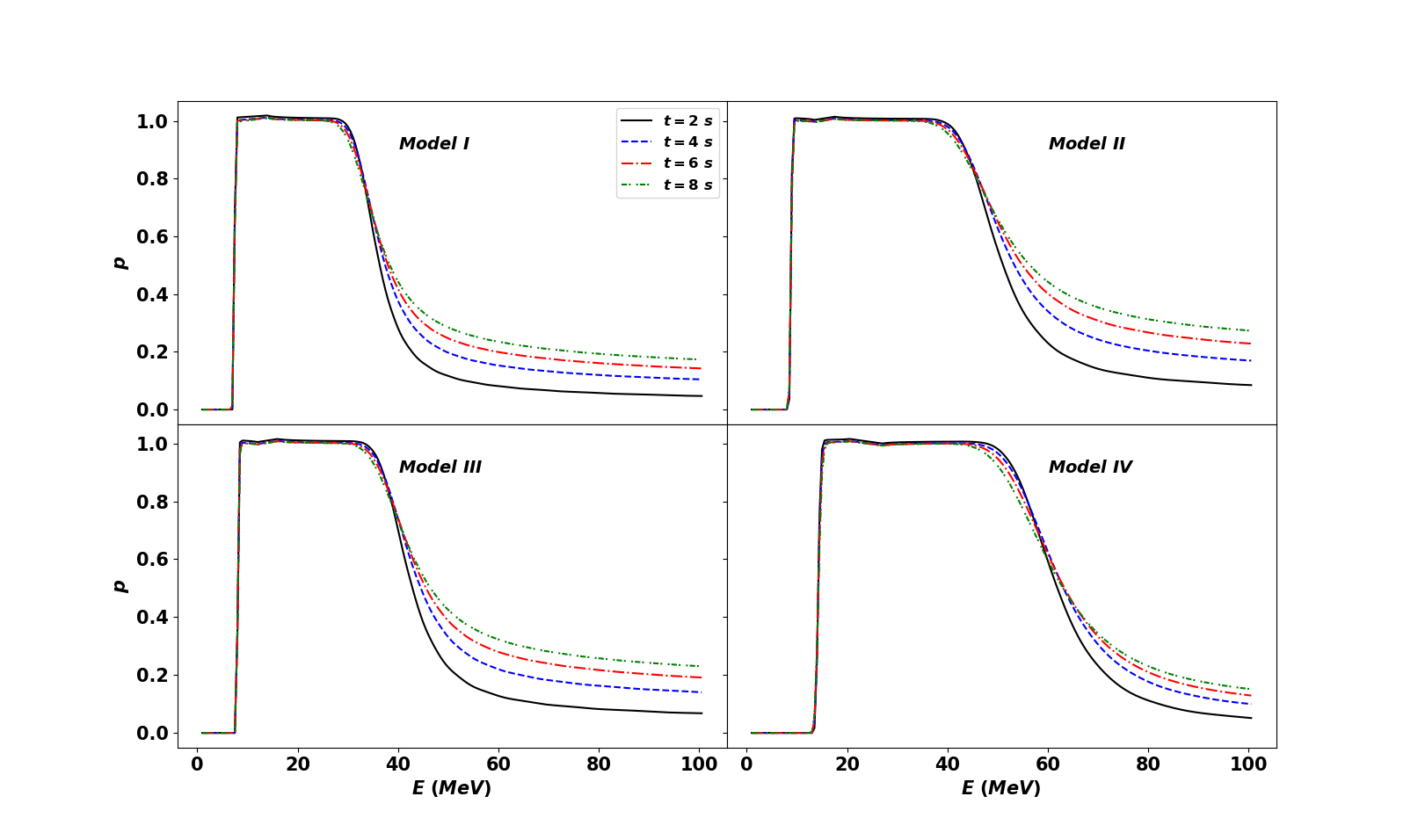}
\caption{The jumping probability $p$ defined in Eq. (\ref{transition
probability}) for models I-IV calculated as a function of energy at even
seconds. At earlier times, the spectral swap is closer to being sharp. But at
later times, the splits become less and less sharp in the high energy region. }
\label{p_figure}
\end{figure}

In the IH case, spectral splits are not sharp in the high energy region of the
spectrum. At high energies, the jumping probability $p$ takes values between $0$
and $1$ as shown in Fig. \ref{p_figure}. In this figure, we show the jumping
probability $p$ for models I-IV as a function of energy at even times. In
earlier times, the spectral swap is closer to being sharp with $p\approx 0.1$ in
the high energy region. But at later times, the splits become less and less
sharp.

\begin{figure}
\begin{center}
\includegraphics[clip=true,trim={0.5cm 0cm 0.5cm 1cm},width=\columnwidth]{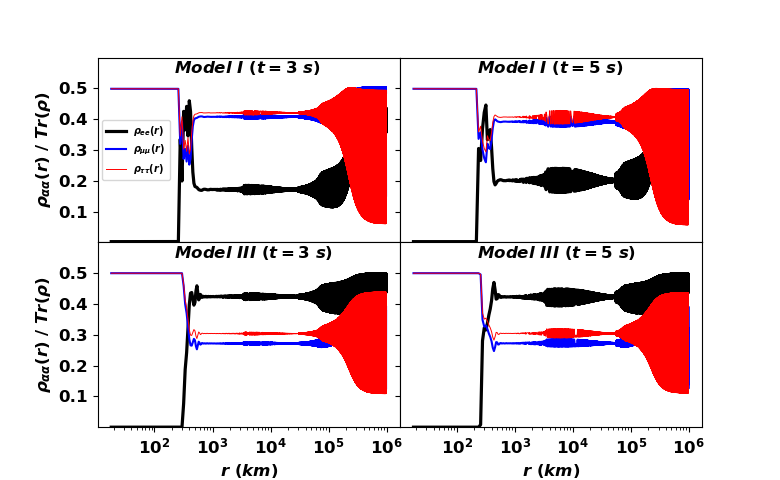}
\caption{The oscillations of the diagonal elements of the density
operator in flavor basis due to the partial adiabaticity of collective
oscillations. These examples show a $45$ MeV neutrino in IH for models I and III
at $3$ s and $5$ s. For other models, energies, and times, the behavior is
similar.} 
\label{oscillations}
\end{center}
\end{figure}

The departure from a sharp spectral split manifests itself most notably in the
oscillations of the diagonal density operator components in flavor basis. This
can be seen by taking the expectation values of $\hat\rho(r)$ given in Eq.
(\ref{final rho in matter}) between the flavor states $\nu_\alpha$, which leads
to
\begin{align}
\label{final rho ee}
&\rho_{\alpha\alpha}(r)=
\left((1-p)\;\rho_{\mu\mu}(R)+p\;\rho_{ee}(R)\right) |\langle \nu_\alpha|r_1\rangle|^2 \nonumber \\
& + \rho_{\mu\mu}(R)|\langle \nu_\alpha|r_1\rangle|^2 \\
& + \left(p\;\rho_{\mu\mu}(R)+(1-p)\;\rho_{ee}(R)\right)|\langle\nu_\alpha|r_3\rangle|^2 \nonumber \\
& + 2\sqrt{p(1-p)}\left(\rho_{ee}(R)-\rho_{\mu\mu}(R)\right)|\langle\nu_\alpha|r_1\rangle
\langle r_3|\nu_\alpha\rangle|\cos\delta(r). \nonumber
\end{align}
As the neutrinos move through the star, the projections of matter eigenstates on
flavor eigenstates (i.e, $\langle \nu_\alpha|r_i\rangle$) change slowly. As a
result, $\rho_{\alpha\alpha}(r)$ has a smooth variation described by the first
three terms of the equation above. However, the last term gives rise to fast
oscillations if $p$ is different from $0$ or $1$. An example of this behavior is
shown in Fig. \ref{oscillations} where we plot $\rho_{\alpha\alpha}(r)$ for a
neutrino with $45$ MeV energy in IH for models I and III at $t=3$ s and $t=5$s.
Notice that the amplitude of the oscillations grows in each case as the neutrino
propagates. This is due to the $\langle \nu_\alpha|r_1\rangle\langle
r_3|\nu_\alpha\rangle$ term in the last line of Eq.  (\ref{final rho ee}). In
the inner regions where the electron density is high, $|\nu_e\rangle$ is
practically the heaviest matter eigenstate. In other words, we have $\langle
\nu_e|r_3\rangle\approx 1$, $\langle \nu_e|r_1\rangle\approx 0$, and $\langle
\nu_{\mu,\tau}|r_3\rangle\approx 0$.  As a result, the oscillation amplitudes do
not grow until the density drops and $\nu_e$ starts to have a projection on
lighter matter eigenstates.  These oscillations are akin to the \emph{phase
effects} first discussed in Ref. \cite{Dasgupta:2005wn}.

\subsection{MSW Resonances}

\begin{figure}
\begin{center}
\includegraphics[clip=true,trim={0.8cm 0cm 1cm 1cm},width=\columnwidth]{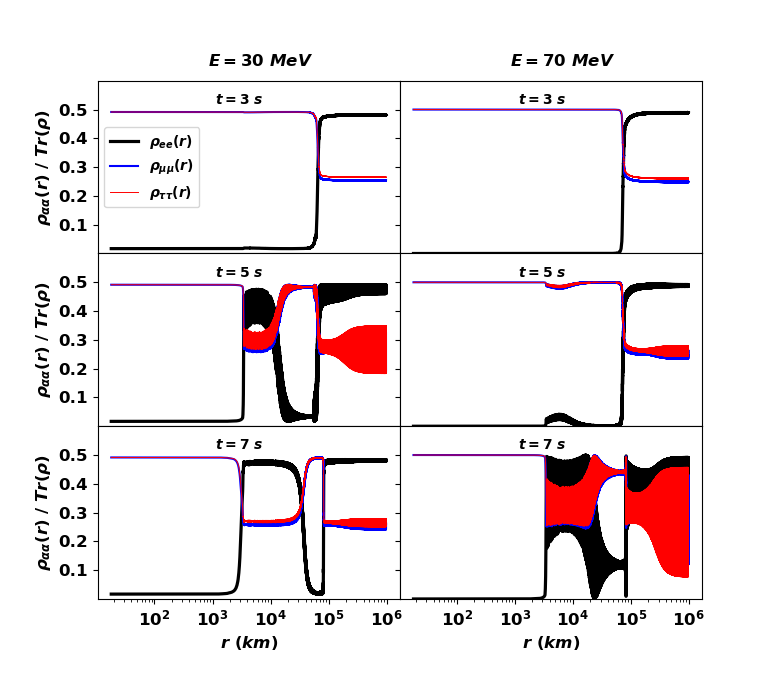}
\caption{The evolution of $30$ MeV (left panels) and $70$ MeV
(right panels) neutrinos in the case of NH for model I at $3$ s, $5$ s, and $7$
s. At earlier times, neutrinos go through one adiabatic high resonance. At later
times, they go through three high resonances, and the adiabaticity is temporarily
broken for each energy.}
\label{msw}
\end{center}
\end{figure}
Neutrino-neutrino interactions become negligible after a few hundred kilometers
from the center. In our simulations, we manually turn off the neutrino-neutrino
interaction term at $1000$ km, after which the Hamiltonian contains only the
vacuum oscillations and the effects of the other background particles. The
evolution of neutrinos under these conditions is well understood in terms of the
MSW resonances \cite{Wolfenstein:1977ue, Mikheev:1986wj}. As long as the density
profile changes slowly in comparison to the neutrino oscillation wavelengths,
the flavor evolution is adiabatic, i.e., the dynamical evolution of each matter
eigenstate follows its own slow change with the background density
\cite{Kuo:1989qe}. At the resonant densities, the flavor content of matter
eigenstates change very fast and the adiabaticity is easiest to break.

In the case of NH, neutrinos first go through the high resonance. The horizontal
lines in Fig. \ref{fig:density} show the high resonance densities for $30$ MeV
and $70$ MeV neutrinos for illustration. A resonance occurs where they cross the
supernova density profile. At earlier times, neutrinos experience only one high
resonance, but at later times multiple high resonances occur due to the low
density region between the front shock and the reverse shock. The resulting
flavor evolutions are shown in Fig. \ref{msw} for $30$ MeV (left panels) and
$70$ MeV (right panels) neutrinos at $t=3$ s, (upper panels) $t=5$ s (middle
panels), and  $t=7$ s (lower panels) for model I in the case of NH. The
neutrinos are not affected by collective oscillations in the inner regions. At
$t=3$ s, both neutrinos experience only one adiabatic high resonance
transformation around $10^5$ km. At $t=5$ s, the $30$ MeV neutrino goes through
three resonances. The effects of the first and second resonances mostly cancel
each other. In other words, after the second resonance, $\rho_{\alpha\alpha}(r)$
come close to their pre-resonance values. However, the adiabaticity is broken as
evidenced by the presence of the oscillations.\footnote{The argument is similar
to the one presented above: When the adiabaticity is partially (but not
completely) broken, an oscillation term appears an in the last line of Eq.
(\ref{final rho ee}).} At $t=5$ s, the density is close but not yet equal to the
resonance value for $70$ MeV neutrino in between the front and the reverse
shock. This creates the small bumps in middle right panel. But, other than that,
this neutrino goes through its regular adiabatic high resonance as it did at
$t=3$ s. At $t=7$ s, $30$ MeV neutrino still goes through three resonances but
the first two resonances now completely cancel each other. The adiabaticity is
also restored. For this neutrino, $\rho_{\alpha\alpha}$ values on the surface of
the star are almost the same at $t=3$ s and at $t=7$ s. The $70$ MeV neutrino
also starts to go through three resonances at $t=7$ s and its adiabaticity
temporarily is broken.

This example illustrates the general behavior that we observe in all of our
calculations in NH. The arrival of the shock wave to high MSW resonance region
has only a limited effect on the neutrino survival probabilities. Instead of one
high resonance, neutrinos go through three high resonances but, the first two
resonances mostly cancel each other. Also the adiabaticity is initially violated
but later restored. See Ref. \cite{Tomas:2004gr} for a more in-depth discussion
of the effects of the shock wave on the MSW resonances.

The low MSW resonance lies closer to the surface of the star, and it is
experienced by neutrinos in both NH and IH. Fig. \ref{fig:density} shows the low
resonance density value for a $45$ MeV neutrino for illustration. In IH, its
effect can be seen in Fig. \ref{oscillations} at around $5\times 10^5$ km where
$\rho_{ee}$ starts to increase. We do not see the effect of low resonance in
Fig. \ref{msw} because in NH it mainly causes transformations between $\nu_\mu$
and $\nu_\tau$, which already evolved similarly up to that point. The low
resonance is always adiabatic. The shock wave arrives at this region later than
the $9$ s mark, by which time the neutrino luminosity is already low, and we
stop our calculations.

\subsection{Approximate Degeneracy Between NH and IH}

A degeneracy was reported in Ref. \cite{Vaananen:2011bf} between the total event
counts in NH and IH cases. We find this to be approximately true. The source of
this approximate degeneracy is easy to understand in light of the above
discussion. 

Since the low resonance is always adiabatic, Eq. (\ref{final rho in matter}) is
valid through the surface of the star for IH with only the definitions of the
matter eigenstates changing with distance. For NH, this is also the case  if we
assume that the first and the second high resonances perfectly cancel each other
at later times and ignore the temporary violations of adiabaticity.  Once the
neutrinos reach the vacuum, we have
\begin{equation}
\label{surface eigenbasis}
\begin{split}
|r_1\rangle  & = |\nu_1\rangle \quad |r_2\rangle  = |\nu_2\rangle \quad
|r_3\rangle  = |\nu_3\rangle \quad \mbox{for NH,} \\
|r_1\rangle  & = |\nu_3\rangle \quad |r_2\rangle  = |\nu_1\rangle \quad
|r_3\rangle  = |\nu_2\rangle \quad \mbox{for IH.}
\end{split}
\end{equation}
Substituting this into Eq. (\ref{final rho in matter}), we find that 
their density operator is given by
\begin{align}
\label{final rho NH}
\hat \rho&(r)=
\left((1-p)\;\rho_{\mu\mu}(R)+p\;\rho_{ee}(R)\right) |\nu_1\rangle\langle \nu_1| \nonumber \\
& + \rho_{\mu\mu}(R)|\nu_2\rangle\langle \nu_2| \\
& + \left(p\;\rho_{\mu\mu}(R)+(1-p)\;\rho_{ee}(R)\right)|\nu_3\rangle\langle \nu_3| \nonumber \\
& +
\sqrt{p(1-p)}\left(\rho_{ee}(R)e^{i\delta(r)}-\rho_{\mu\mu}(R)e^{-i\delta(r)}\right)|\nu_1 \rangle\langle \nu_3| \nonumber\\
& + \mbox{h.c.} \nonumber
\end{align}
for NH, and by
\begin{align}
\label{final rho IH}
\hat \rho&(r)=
\left((1-p)\;\rho_{\mu\mu}(R)+p\;\rho_{ee}(R)\right) |\nu_3\rangle\langle \nu_3| \nonumber \\
& + \rho_{\mu\mu}(R)|\nu_1\rangle\langle \nu_1| \\
& + \left(p\;\rho_{\mu\mu}(R)+(1-p)\;\rho_{ee}(R)\right)|\nu_2\rangle\langle \nu_2| \nonumber \\
& + \sqrt{p(1-p)}\left(\rho_{ee}(R)e^{i\delta(r)}-\rho_{\mu\mu}(R)e^{-i\delta(r)}\right)|\nu_3 \rangle\langle \nu_2| \nonumber\\
& + \mbox{h.c.} \nonumber
\end{align}
for IH as they leave the star.

After the neutrinos leave the star, they travel a long distance to reach the
Earth.  Over such distances, one should take \emph{neutrino decoherence} into
account, which is the fact that the mass eigenstates traveling with different
speeds open up a gap between them. For $r>r_{\mbox{\tiny coh}}$, the gap becomes
larger than their wavepacket size, and they cease to overlap. After that, the
off-diagonal terms of the density matrix in the mass basis decrease with
$e^{-(r/r_{\mbox{\tiny coh}})^2}$. For supernova neutrinos, the coherence length
is of the order of a fraction of a parsec. By the time the neutrinos travel $10$
kpc, their density operator is given only by the diagonal components of those
given in Eqs.  (\ref{final rho NH}) and (\ref{final rho IH}). To calculate the
detector response, we only need the $\nu_e$ component of the density operator,
which is given by
\begin{align}
\label{earth rho NH}
\hat \rho_{ee}(d)=&
\left((1-p)\;\rho_{\mu\mu}(R)+p\;\rho_{ee}(R)\right) U_{e1}^2
 + \rho_{\mu\mu}(R) U_{e2}^2 \nonumber \\
& + \left(p\;\rho_{\mu\mu}(R)+(1-p)\;\rho_{ee}(R)\right) U_{e3}^2
\end{align}
for NH and by
\begin{align}
\label{earth rho IH}
\hat \rho_{ee}(d)=&
\left((1-p)\;\rho_{\mu\mu}(R)+p\;\rho_{ee}(R)\right) U_{e3}^2
+ \rho_{\mu\mu}(R) U_{e1}^2 \nonumber\\
& + \left(p\;\rho_{\mu\mu}(R)+(1-p)\;\rho_{ee}(R)\right) U_{e2}^2
\end{align}
for IH, where $d$ is the distance between the Earth and the supernova.

Let us assume that sharp spectral splits develop as in Eqs. (\ref{p NH}) and
(\ref{p IH}). The NH split energy $E_{\mbox{\tiny NH}}$ is always lower than
both of the IH split energies $E_{\mbox{\tiny IH}}$ and $E_{\mbox{\tiny IH}}'$.
Therefore, in the energy region $E_{\mbox{\tiny IH}}<E<E_{\mbox{\tiny IH}}'$
swapped by the collective oscillations, we have
\begin{align}
\label{degenerate rho} 
\hat \rho_{ee}(d)\approx\rho_{ee}(R)U_{e3}^2 + \rho_{\mu\mu}(R) (1-U_{e3}^2)  
\end{align}
for both NH and IH. This formula can be found by substituting $p=0$ in Eq.
(\ref{earth rho NH}) and by substituting $p=1$ in Eq. (\ref{earth rho IH}). One
also also needs to use the fact that $U_{e1}^2+U_{e2}^2+U_{e3}^2=1$. This leads
to a near degeneracy between the NH and IH event rates because HALO is most
sensitive in the energy region which is typically swapped by collective
oscillations in the IH case. However the degeneracy is broken by several
factors. This includes the departure from sharp spectral splits, and the
reactions caused by neutrinos with energy higher than $E_{\mbox{\tiny IH}}'$. 

\section{Reaction Rates}
\begin{figure*}
\begin{center}
\includegraphics[clip=true,trim={0cm 0cm 0cm 0cm},width=\textwidth]{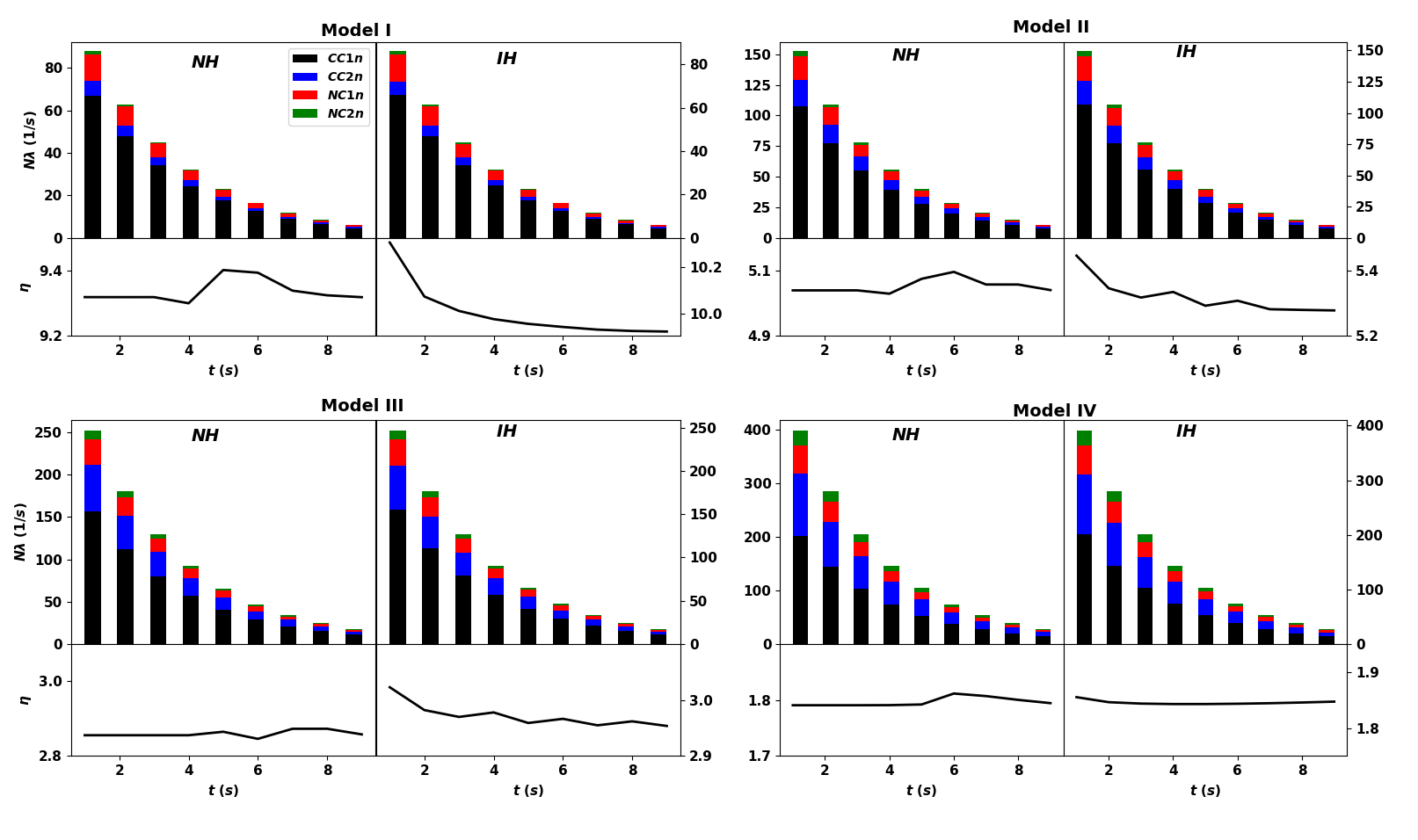}
\caption{Calculated reaction rates as functions of post-bounce
time for models I-IV. Upper panels show the rates of CC1n, CC2n, NC1n, and NC2n
reactions as indicated in the first panel. Lower panels show the ratio of 1n to
2n reaction rates. The left panels correspond to NH, and the right panels
correspond to IH.}
\label{reaction_rates}
\end{center}
\end{figure*}

The flux of neutrinos of type $\nu_\alpha$ with energy between $E$ and $E+dE$ at
Earth is given by\footnote{This flux involves neutrinos going in all directions,
not just the radial ones. To obtain the flux of only those neutrinos traveling
radially, one should further divide this flux with another factor of $\pi$ as
shown in Ref. \cite{Duan:2006an}.} $(\nicefrac{1}{4\pi d^2})
\rho_{\alpha\alpha}(d) dE$. Therefore, the number of a particular kind of
reaction per unit time per target nucleus is equal to 
\begin{equation}
\label{rate1}
\lambda(t)=\frac{1}{4\pi d^2}\int_{E_{\mbox{\tiny th}}}^\infty 
\rho_{\alpha\alpha}(d) \sigma(E) dE. 
\end{equation}
Here $\sigma(E)$ and $E_{\mbox{\tiny th}}$ denote the cross section and the
threshold energy of the reaction, respectively. $\rho_{\alpha\alpha}(d)$ depends
on the post-bounce time $t$ and the neutrino energy $E$ implicitly. If there
are $N$ target nuclei in the detector, the rate of this particular reaction is
given by $N\lambda(t)$. For a detector of mass $M$, this is equal to 
\begin{equation}
\label{rate2}
\begin{split}
N\lambda(t)=&241 \mbox{s}^{-1}
\left(\frac{M}{1 \mbox{kton}}\right)
\left(\frac{10 \mbox{kpc}}{d}\right)^2
\\
&\int_{E_{\mbox{\tiny th}}}^\infty 
\left(\frac{\sigma(E)}{10^{-40}\,\mbox{cm}^2}\right) 
\left(\frac{\rho_{\alpha\alpha}(d)}{10^{58}\,
\mbox{s}^{-1}\mbox{MeV}^{-1}} \right)\frac{dE}{\mbox{MeV}}. 
\end{split}
\end{equation}

In calculating the reaction rates, we use the cross sections provided in Ref.
\cite{Engel:2002hg} with necessary interpolations to our energy bins. The
threshold energies of the reactions are calculated from measured masses and
given in Table \ref{parameters}.  The density matrix elements
$\rho_{\alpha\alpha}(d)$ are calculated by numerically evolving Eq. (\ref{equ1})
through the relevant density profile at each second.

In HALO, the outgoing electron is not detected \cite{Schumaker:2010ata}. For
this reason, CC and NC reactions cannot be distinguished from each other. But 1n
and 2n events can be discriminated. Neither the detection nor the discrimination
of 1n and 2n events is 100\% efficient. But in our calculations, we assume
perfect efficiency.

Fig. \ref{reaction_rates} shows the time dependent reaction rates that we
calculate using Eq. (\ref{rate2}) per $1$ kt detector mass for a supernova which
is $10$ kpc away from the Earth. The left panels correspond to NH and the right
panels correspond to IH. The upper panels show the individual rates of CC1n,
CC2n, NC1n, and NC2n reactions. These reaction rates are also shown in Table
\ref{table NH} for NH and in Table \ref{table IH} for IH. Note that the reaction
rates we show in the tables are rounded to integer numbers whereas those plotted
in Fig. \ref{reaction_rates} are not. The lower panels of Fig.
\ref{reaction_rates} show the ratio of the total rate of 1n reactions to the
total rate of 2n reactions, i.e.,  
\begin{equation}
\label{r}
\eta(t)=\frac 
{\lambda_{\mbox{\tiny CC1n}}(t)+\lambda_{\mbox{\tiny NC1n}}(t)}
{\lambda_{\mbox{\tiny CC2n}}+\lambda_{\mbox{\tiny NC2n}}(t)}.
\end{equation}
Here, indices are used to refer to the rates of individual reactions. This ratio
is independent from the detector size and from the distance of the supernova. If
the effects of the shock wave and the changing character of collective
oscillations are not considered, $\eta$ would be independent of time.  But even
when these effects are taken into account, our results show that $\eta$ depends
only very weakly on time. For the models that we consider, we find that $\eta$
changes by no more than a few percent with time for both NH and IH. The most
substantial change is observed in model I in the case of IH where $\eta$ drops
from $10.3$ to $9.9$, which is about a $4\%$ change. This tells us that the time
dependent features that we consider here, i.e., the loss of sharpness in
spectral splits with the decreasing neutrino luminosity, and the passage of the
shock wave from the MSW resonance region, are likely to be lost within the error
bars unless the statistics of the experiment is significantly improved. 

As discussed above, there is a near degeneracy between NH and IH scenarios. In
most cases, NH and IH results differ at most by a few events per second. But we
find that, in each case, $\eta$ is slightly larger for IH. 

As we go from model I to model IV, all reaction rates increase. This is expected
because $\langle E_{\nu_x}\rangle$ increases from model I to model IV. The
energetic  $\nu_\mu$ - $\nu_\tau$ neutrinos are converted to $\nu_e$ by
collective oscillations in the case of IH and by the high resonances in the case
of NH. But we also see that $\eta$ decreases as we go from model I to model IV.
This tells us that the 2n event rates  increase to a greater extent with
$\langle E_{\nu_x}\rangle$. 

\begin{table}
\begin{center}
\begin{tabular}{| c | c | c | c | c | c | c | c | c | c | c |}
\hline 
&time (s)&\; $1$ \; & \; $2$ \; & \; $3$ \; & \; $4$ \; & \; $5$ \; & \;
$6$ \; & \; $7$ \; & \; $8$ \; & \; $9$ \; \\ 
\hline 
\multirow{4}{*}{\rotatebox[origin=c]{90}{Model I}} 
& CC1n & 67 &  48 &  34 &  24 &  18 &  13 &   9 &   6 &   5 \\
& NC1n & 12 &   9 &   6 &   5 &   3 &   2 &   2 &   1 &   1 \\
& CC2n & 7 &   5 &   4 &   3 &   2 &   1 &   1 &   1 &   0 \\
& NC2n & 1 &   1 &   1 &   0 &   0 &   0 &   0 &   0 &   0  \\
\hline 
\multirow{4}{*}{\rotatebox[origin=c]{90}{Model II}} 
& CC1n & 108 &   77 &   55 &   40 &   28 &   20 &   15 &   10 &    7\\
& NC1n & 19 &   14 &   10 &    7 &    5 &    4 &    3 &    2 &    1\\
& CC2n & 21 &   15 &   11 &    8 &    6 &    4 &    3 &    2 &    1\\
& NC2n & 4 &    3 &    2 &    1 &    1 &    1 &    1 &    0 &    0 \\
\hline 
\multirow{4}{*}{\rotatebox[origin=c]{90}{Model III}} 
& CC1n & 156 &  112 &   80 &   58 &   41 &   29 &   21 &   15 &   11 \\
& NC1n & 30 &   22 &   16 &   11 &    8 &    6 &    4 &    3 &    2 \\
& CC2n & 55 &   39 &   28 &   20 &   14 &   10 &    7 &    5 &    4 \\
& NC2n & 10 &    7 &    5 &    4 &    3 &    2 &    1 &    1 &    1 \\
\hline 
\multirow{4}{*}{\rotatebox[origin=c]{90}{Model IV}} 
& CC1n &  202 &  145 &  104 &   74 &   53 &   38 &   27 &   20 &   14 \\
& NC1n &53 &   38 &   27 &   20 &   14 &   10 &    7 &    5 &    4 \\
& CC2n &116 &   83 &   59 &   43 &   30 &   21 &   15 &   11 &    8 \\
& NC2n &27 &   19 &   14 &   10 &    7 &    5 &    4 &    3 &    2 \\
\hline 
\end{tabular}
\end{center}
\caption{Calculated reaction rates (rounded to integers), in units of $\text{s}^{-1}$, for the case of NH as
a function of time.}
\label{table NH}
\end{table}
\begin{table}
\begin{center}
\begin{tabular}{| c | c | c | c | c | c | c | c | c | c | c |}
\hline 
&time (s)&\; $1$ \; & \; $2$ \; & \; $3$ \; & \; $4$ \; & \; $5$ \; & \;
$6$ \; & \; $7$ \; & \; $8$ \; & \; $9$ \; \\ 
\hline 
\multirow{4}{*}{\rotatebox[origin=c]{90}{Model I}} 
& CC1n & 66 &  47 &  34 &  24 &  17 &  12 &   9 &   6 &   5\\
& NC1n &12 &   9 &   6 &   5 &   3 &   2 &   2 &   1 &   1\\
& CC2n &6 &   5 &   3 &   2 &   2 &   1 &   1 &   1 &   0\\
& NC2n &1 &   1 &   1 &   0 &   0 &   0 &   0 &   0 &   0\\
\hline 
\multirow{4}{*}{\rotatebox[origin=c]{90}{Model II}} 
& CC1n & 107 &   76 &   55 &   39 &   28 &   20 &   14 &   10 &    7\\
& NC1n &19 &   14 &   10 &    7 &    5 &    4 &    3 &    2 &    1\\
& CC2n &19 &   14 &   10 &    7 &    5 &    4 &    3 &    2 &    1\\
& NC2n &4 &    3 &    2 &    1 &    1 &    1 &    1 &    0 &    0\\
\hline 
\multirow{4}{*}{\rotatebox[origin=c]{90}{Model III}} 
& CC1n &155 &  110 &   79 &   57 &   41 &   29 &   21 &   15 &   11\\
& NC1n &30 &   22 &   16 &   11 &    8 &    6 &    4 &    3 &    2\\
& CC2n &51 &   37 &   27 &   19 &   14 &   10 &    7 &    5 &    4\\
& NC2n &10 &    7 &    5 &    4 &    3 &    2 &    1 &    1 &    1 \\
\hline 
\multirow{4}{*}{\rotatebox[origin=c]{90}{Model IV}} 
& CC1n &201 &  143 &  103 &   74 &   53 &   38 &   27 &   19 &   14\\
& NC1n &53 &   38 &   27 &   20 &   14 &   10 &    7 &    5 &    4\\
& CC2n &110 &   79 &   57 &   41 &   29 &   21 &   15 &   11 &    8\\
& NC2n &27 &   19 &   14 &   10 &    7 &    5 &    4 &    3 &    2 \\
\hline 
\end{tabular}
\end{center}
\caption{Calculated reaction rates (rounded to integers), in units of $\text{s}^{-1}$, for the case of IH as a
function of time.}
\label{table IH}
\end{table}

\section{Conclusions}

In this paper, we calculated the event rates in a lead-based detector due to a
galactic core-collapse supernova. We paid particular attention to the time
dependence of the reaction rates due to the flavor evolution of neutrinos
through time dependent conditions in the supernova.  In particular, we focused
on the changing character of collective neutrino oscillations due to the
decreasing neutrino luminosity, and the propagation of the shock wave through
the MSW region. 

For this purpose, we formed a one-dimensional supernova model by superimposing a
parametric shock wave on a progenitor density distribution which models SN1987A.
We considered four different models in this setting. These models have the same
neutrino luminosities (which is consistent with SN1987A) but the initial
neutrino energy distributions are different. In models I-III, we kept the
$\nu_e$ and $\bar\nu_e$ distributions fixed while shifting $\nu_x$ distributions
to higher energies. In model IV we used higher average energies for all flavors,
which are far-fetched but not completely ruled out \cite{Mezzacappa:2000jb,
Mathews:2014qba}. 

For all models, we find that the sharpness of spectral splits decrease with time
in the IH case. Since neutrinos do not go through high MSW resonance, this is
the only source of time dependence in IH. However, we find that the resulting
effect on reaction rates is limited because loss of sharpness becomes noticeable
only when the neutrino luminosity drops considerably.
 
In the case of NH, neutrinos are not affected by collective oscillations but go
though the high MSW resonance. In this case, the passage of the shock wave
through the high MSW region is the only source of time dependence. Initially
neutrinos experience only one high resonance but after the shock wave passes
through this region, they start going through three high resonances. We find
that the effects of the first two resonances mostly cancel each other, and the
adiabaticity is only temporarily lost. For this reason, the resulting effect on
the reaction rates is also limited. 

All reaction rates decrease roughly exponentially as the neutrino luminosity
drops. The ratio of 1n to 2n event rates is the best parameter to work with
because it is independent of this overall decrease. We find that this ratio
changes slightly with time due to the above-mentioned effects. In the case of
IH, it slightly decreases with time. In the case of NH, it is initially
constant, but later it changes as the shock wave passes through the high MSW
resonance region.  This effect is similar to the one observed in Ref.
\cite{Gava:2009pj} for the electron antineutrino signal in water-Cherenkov
detectors in the case of IH.  However, in all the models that we looked at, the
change is limited to a few percent. Therefore, it is likely to be lost within
the error bars, or within the time dependence resulting the evolution of the
proto-neutron star itself. 

The evolution of the proto-neutron star is something that we intentionally left
out in this paper. Our purpose was to isolate the time dependence resulting from
the dynamical flavor evolution of neutrinos outside the proto-neutron star.  We
also left out the multiangle nature of the collective neutrino oscillations. The
multiangle effects delay the appearance of collective effects. They may also
introduce angular decoherence and cause spectral splits to be less sharp, or
even wash them out completely. For the type of initial spectra that we consider,
in which $\nu_e$ and $\nu_\mu-\nu_\tau$ distributions cross each other only
once, these effects appear to be minimal \cite{Mirizzi:2010uz}. But in fact, the
nature of the multiangle collective oscillations remains to be fully understood.
See, e.g. Refs. \cite{Chakraborty:2014nma, Chakraborty:2016yeg,
Morinaga:2019wsv}. At this point, we believe that their inclusion is not likely
to change the main results of this paper.

\vspace*{3mm} \noindent 
Y.P. thanks the APS Gordon and Betty Moore Foundation for their Visitor
Award, which provided funding for a visit to University of Wisconsin-Madison
(Grant no. GBMF6210).
B. E. acknowledges the 2214A travel fellowship from the Scientific and Technological
Research Council of Turkey (T{\"{U}}B{\.{I}}TAK). Y.P. and B.E. thank to 
the University of Wisconsin-Madison their hospitality. 
This work was supported in part 
by T{\"{U}}B{\.{I}}TAK under Project No. 117F327. 
The work of A.~V.~P. was supported in part by the NSF (Grant no. PHY-1630782)
and the Heising-Simons Foundation (2017-228), and in part by the U.S. Department
of Energy under contract number DE-AC02-76SF00515.

\bibliography{bibyamac}

\end{document}